\documentclass[12pt,reqno]{amsart}
\makeatletter
\def\subsection{\@startsection{subsection}{2}%
  \z@{.5\linespacing\@plus.7\linespacing}{.5\linespacing}%
  {\normalfont\bfseries}}
\makeatother
\allowdisplaybreaks[4]

\usepackage [latin1]{inputenc}
\usepackage{graphicx}
\usepackage{amssymb}
\usepackage{color}
\usepackage{epstopdf}
\usepackage{amsthm,amsmath,amssymb}
\usepackage{mathrsfs}
\usepackage{cite}
\usepackage{subeqnarray}
\usepackage{cases}
\usepackage{authblk}
\usepackage{hyperref}

{   \theoremstyle{plain}

}
\newtheorem{thm}{Theorem}[section]

\begin{document}
\begin{center}
    {\large \sc \bf  On one typical Einstein-Weyl equation:
    inverse spectral transform for the
    Cauchy problem, longtime behaviour
    of the solutions and implicit solutions}

\vskip 20pt
    {Ge Yi, Zikai Chen, Kelei Tian, Ying Xu$^*$}	
{\large }

    \vskip 20pt

    {\it
    School of Mathematics, Hefei University of Technology, Hefei 230601, China
     }

    \bigskip

    $^*$ Corresponding author:{\tt xuying@hfut.edu.cn }
    
    \bigskip

    {\today}

    \end{center}
    \bigskip
    \bigskip
    \textbf{Abstract:} In this paper, we study one typical Einstein-Weyl  equation. It arises from Ferapontov and Kruglikov's investigation on the integrability of several dispersionless partial differential equations and the geometry of their formal linearizations. First, by using  Manakov-Santini IST (inverse spectral transform) method, we investigate its Cauchy problem including the direct problem from the initial data,  the time evolution of the scattering and spectral data, and the inverse problem. Second, based on the nonlinear RH (Riemann-Hilbert) dressing, the longtime behaviour of the solutions is constructed.
    In addition, some implicit solutions are presented.
    \bigskip
    \bigskip
\section{\sc \bf Introduction}
 Einstein-Weyl geometry, introduced by Weyl in an attempt to unify  electromagnetic and gravitational fields within the framework of geometry\cite{RN28},
has received extensive attention in recent decades. In 1990, Ward
demonstrated the existence of  a special class of Einstein-Weyl
spaces corresponding to solutions of the Toda field equation for the group SU($\infty$), which provides a perspective for studying conformal field theory\cite{RN1}.
In 1993, Tod investigated various differential
geometric properties of three-dimensional Einstein-Weyl
spaces, giving some examples of such spaces and contributing to the development of their general theory\cite{RN2}.
In 2001, Dunajski, Mason, and Tod explored the relationship between  Einstein-Weyl geometry and the dKP (dispersionless Kadomtsev-Petviashvili) equation. They showed that Einstein-Weyl equations in (2+1)-dimensional space contain the dKP equation as a special case\cite{RN3}. In 2002,  Dunajski and Tod showed how the solutions of the dKP equation can be used to construct certain  three-dimensional Einstein-Weyl structures\cite{RN4}. In 2004, Dunajski demonstrated that the solutions of  a pair of quasi-linear partial differential equations can be used to construct Lorentzian Einstein-Weyl structures in (2+1)-dimensional space \cite{RN5}.\par
The dispersionless integrable systems, which have attracted significant focus in mathematical physics and theoretical physics, are an important class of integrable systems.
They are often equivalent to the commutation
condition of vector field Lax pairs. In 1991,  Krichever pointed out that the dispersionless (or semiclassical) limits of the classical KP (Kadomtsev-Petviashvili) hierarchy is the dKP hierarchy\cite{RN6}.
In 1992, motivated by  Krichever's works, Takasaki and Takebe proposed the SDiff(2) KP hierarchy related to the group of
area-preserving diffeomorphisms on a cylinder\cite{RN24}.
In 1997, Guha and Takasaki  established a dictionary between the twistor geometry and the
method of RH problem for both the dKP and dToda (dispersionless Toda) hierarchies\cite{RN7}.
In 2011, Takasaki and Nakatsu showed that the solution of a Riemann-Hilbert problem associated with thermodynamic limit of random partition models is identified with a special solution of the dToda  hierarchy\cite{RN25}.
In 2012, Takasaki studied geometric aspects of two particular types of finite-variable reductions in the dToda hierarchy\cite{RN9}.
In 2015, Takasaki introduced the orbifold generalizations of the ordinary and modified melting crystal models, and pointed out that the associated partition functions can be converted to tau functions of the 2D Toda hierarchy\cite{RN26}.
In 2019, Takasaki addressed the problems of quantum spectral curves and 4D limit for the melting crystal model of 5D SUSY U(1) Yang-Mills
theory on $\mathbb{R}^4\times S^1$\cite{RN27}.
In 2014, Ferapontov and Kruglikov showed that the symbols of
the formal linearizations of several classes of second-order dispersionless systems
define conformal structures of Einstein-Weyl geometry in three dimensions
(or self-dual in four dimensions). This implies that
one can see the integrability of these
dispersionless systems from the geometry of their formal
linearizations\cite{RN10}.
In order to read the connection between the integrability of a given partial differential equation and the geometry of its formal linearization, they
studied four classes of dispersionless systems in three dimensions.
An important class among them,
\begin{equation}\label{e0}
    (a(u))_{xx}+(b(u))_{yy}+(c(u))_{tt}+2(p(u))_{xy}+2(q(u))_{xt}+2(r(u))_{yt}=0,
\end{equation}
has been studied in detail\cite{RN10,RN4}.
This class of systems is integrable by the method of hydrodynamic reductions if a symmetric matrix associated with its coefficients  satisfies a particular constraint\cite{RN10,RN11}.
Under this constraint, there are five canonical forms of nonlinear integrable models, including the well-known BF (Boyer-Finley) equation
and  dKP equation.
We concentrate on another model among the five nonlinear integrable models,
\begin{equation}\label{e1}
    (u^2)_{xy}+u_{yy}+2u_{xt}=0,
\end{equation}
where the potential $u(x,y,t)$, the space variables $x$, $y$ and the time variable $t$ are real.
Such dispersionless integrable system can
arise from the commutation of the  Lax pair
\begin{equation}\label{e33}
    \begin{aligned}
        &\hat{L}_1=\partial_y-\lambda\partial_x+2u_x\lambda\partial_\lambda,\\
        &\hat{L}_2=\partial_t+(\frac{1}{2}\lambda^2+u\lambda)\partial_x-(u_x\lambda+u_y+2uu_x)\lambda\partial_\lambda.
    \end{aligned}
\end{equation}\par
For some complicated nonlinear mathematical physics systems, finding solutions is usually very difficult. In integrable systems, the IST method is one significant approach to solving these systems.
It is first introduced
by Gardner, Green, Kruskal and Miura in\cite{RN12}
to solve the Cauchy problem of distinguished KdV
(Korteweg-de Vries) equation. In 1968, Lax summarized their works, making the IST method become a general method of studying classical integrable systems systematically\cite{RN13}. Since then, many important classical integrable systems including their Cauchy problems and soliton solutions have been extensively studied. Around 2006, Manakov and Santini  introduced a novel IST method for solving dispersionless integrable systems\cite{RN14,RN15}. Subsequently, using this novel IST method, they
constructed the formal solution of
the Cauchy problem of a hierarchy of integrable partial differential equations
in (2 + 1) dimensions and the dispersionless 2D Toda equation\cite{RN16,RN17}.\par
Among these dispersionless integrable systems, the dKP equation has undergone relatively systematic study under this novel IST method.
In \cite{RN14,RN18}, Manakov and Santini studied the Cauchy problem, longtime behaviour, wave breaking and implicit solutions of the dKP equation.
In the study of longtime behaviour of the solutions of the dKP equation, based on the space-time region
\begin{equation*}
    \begin{aligned}
        &x=\tilde{x}+v_1t,\;y=v_2t,\;x+\frac{y^2}{4t}=\tilde{x},\\
        &\tilde{x}-2ut,\;v_1,\;v_2=O(1),\;v_2\ne 0,\;t\gg 1,
    \end{aligned}
\end{equation*}
they obtained
\begin{equation*}
    u=\frac{1}{\sqrt{t}}G(x+\frac{y^2}{4t}-2ut,\;\frac{y}{2t})+o(\frac{1}{\sqrt{t}}).
\end{equation*}
From above equation they explored the wave breaking and obtained that the wave breaking of the dKP equation is similar to (1+1)-dimensional Hopf equation, implying that the dKP equation can be viewed
as the integrable universal model in the description of the wave breaking of a localized two-dimensional wave.\par
In this paper, using  Manakov-Santini IST method, we focus on the Cauchy problem and longtime behaviour of the system (\ref{e1}).
The paper is organized as follows.
In section 2, we construct the formal IST to solve
 the Cauchy problem with localized initial data.
We also consider the time evolution of the scattering and spectral data for an associated hierarchy.
In section 3, based on the nonlinear RH dressing, we construct the longtime behaviour of the solutions.
In section 4, we present some implicit solutions .
\section{\sc \bf The Cauchy problem for the equation}
For the evolution systems, both the vanishing-boundary and periodic initial data have strong physical background and have been widely studied in mathematics.
In this section, using Manakov-Santini IST method, we  construct the formal solution  with vanishing-boundary conditions, i.e.,
the equation (\ref{e1}) within the class of rapidly decreasing real potential $u(x,y,t)$,
\begin{equation}\label{e2}
    \begin{aligned}
        u\rightarrow 0, \;(x^2+y^2)\rightarrow\infty.
    \end{aligned}
\end{equation}
\subsection{Jost eigenfunctions and analytic eigenfunctions}\label{subsection1}
Due to the localization (\ref{e2})
of the potential $u$,
the eigenfunction $f(x,y,\lambda)$ of operator $\hat{L}_1$ exhibits the following asymptotics,
\begin{equation*}
    \begin{gathered}
        f(x,y,\lambda)\rightarrow f_{\pm}(\lambda,\gamma),\; y\rightarrow\pm\infty,
    \end{gathered}
\end{equation*}
where $\gamma=x+\lambda y$.\par
The Jost eigenfunctions of operator $\hat{L}_1$, defined on $Im\lambda=0$, play a basic role in the IST theory.
The vector Jost eigenfunctions $\vec{\varphi}_\pm(x,y,\lambda)$ exhibit the following asymptotics,
\begin{equation}\label{e5}
    \vec{\varphi}_\pm(x,y,\lambda)=\left(\begin{gathered}
        \varphi_{\pm1}(x,y,\lambda)\\
        \varphi_{\pm2}(x,y,\lambda)
    \end{gathered}\right)\rightarrow
    \vec{\xi},\;
    y\rightarrow \pm\infty,\;\lambda\in\mathbb{R},
\end{equation}
where
\begin{equation*}
    \vec{\xi}=\left(
    \begin{gathered}
        \lambda\\
        \gamma
    \end{gathered}
    \right),\;\gamma=x+\lambda y.
\end{equation*}\par
There is a connection between
these Jost eigenfunctions and the dynamical system
\begin{equation}\label{e3}
    \frac{\mathrm{d}\vec{v}}{\mathrm{d}y}=\left(
    \begin{gathered}
        2u_x\lambda\\
        -\lambda
    \end{gathered}\right),
\end{equation}
where $\vec{v}(y)=(\lambda(y),x(y))^T$, and $y$ is interpreted as the time variable.\par
Assuming that the potential $u$ is smooth and sufficiently localized function of
$x$ and $y$, it follows from the ODE theory that the solution
\begin{equation*}
    \vec{v}=\left(
    \begin{gathered}
        \Lambda(y,x_0,y_0,\lambda_0)\\
        x(y,x_0,y_0,\lambda_0)
    \end{gathered}\right)
\end{equation*}
of system (\ref{e3}), with the initial condition $(\lambda(y_0),x(y_0))^T=(\lambda_0,x_0)^T$,
exists uniquely. It is globally defined for real $y$, with
the following asymptotic states $\tilde{\lambda}_\pm$, $\tilde{x}_\pm$,
\begin{equation}\label{e4}
    \begin{aligned}
        &\lambda_\pm(x,y,\lambda)\rightarrow\tilde{\lambda}_\pm(x_0,y_0,\lambda_0),\;y\rightarrow\pm\infty,\\
        &x_\pm(x,y,\lambda)\rightarrow -y\tilde{\lambda}_\pm(x_0,y_0,\lambda_0)+\tilde{x}_\pm(x_0,y_0,\lambda_0),\;y\rightarrow\pm\infty.
    \end{aligned}
\end{equation}
The asymptotics (\ref{e4}) imply that when the initial point $(x_0,y_0,\lambda_0)$ moves along the
trajectories, these asymptotic states $\tilde{\lambda}_\pm(x_0,y_0,\lambda_0)$, $\tilde{x}_\pm(x_0,y_0,\lambda_0)$
are constants of motion for the dynamical system (\ref{e3}). Therefore, $\lambda_\pm(x,y,\lambda)$ and $x_\pm(x,y,\lambda)$ are the solutions of the vector field equation
\begin{equation*}
        \hat{L}_1\left(
            \begin{gathered}
                \lambda_\pm(x,y,\lambda)\\
                x_\pm(x,y,\lambda)
            \end{gathered}
        \right)
        =\vec{0}.
\end{equation*}
Due to asymptotics (\ref{e5}) and (\ref{e4}), they coincide with the real Jost eigenfunctions
$\vec{\varphi}_{\pm}(x,y,\lambda)$,
\begin{equation*}
    \begin{gathered}
        \varphi_{\pm1}(x,y,\lambda)=\lambda_\pm(x,y,\lambda),\\
        \varphi_{\pm2}(x,y,\lambda)=x_\pm(x,y,\lambda).
    \end{gathered}
\end{equation*}\par
Usually, the integrable systems are often associated with some RH problems by using the IST method. The vanishing-boundary problems are often associated with the RH problems with jumps along the real axis.
The analytic eigenfunctions that are analytic in the upper and lower halves of complex $\lambda$ plane respectively are concerned particularly.
The analytic eigenfunctions $\vec{\Psi}^{\pm}(x,y,\lambda)$ of operator $\hat{L}_1$
satisfy the asymptotics
\begin{equation*}
    \vec{\Psi}^{\pm}(x,y,\lambda)\rightarrow\vec{\xi},\;(x^2+y^2)\rightarrow\infty,
\end{equation*}
where $\vec{\Psi}^{+}$ is analytic in $Im\lambda>0$ and continuous in
$Im\lambda\ge 0$, while $\vec{\Psi}^{-}$ is analytic in $Im\lambda<0$ and continuous in
$Im\lambda\le 0$.\par
These analytic eigenfunctions can be characterized by the  integral equations
\begin{equation*}
    \begin{aligned}
        \vec{\Psi}^{\pm}(x,y,\lambda)=\vec{\xi}+
        \int_{\mathbb{R}^2}\mathrm{d}x'dy'G^{\pm}(x-x',y-y',\lambda)
        (-2u_{x'}(x', y')\lambda{\vec{\Psi}^{\pm}_\lambda}(x', y',\lambda)),
    \end{aligned}
\end{equation*}
where
\begin{equation}\label{e6}
    G^{\pm}(x,y,\lambda)=\mp\frac{1}{2\pi i[x+(\lambda\pm i\epsilon)y]}
\end{equation}
are the analytic Green's functions, and the symbol $\int_{\mathbb{R}^2}$ represents the double integral over $\mathbb{R}^2$.\par
The analyticity properties of $G^{\pm}(x,y,\lambda)$ guarantee the analyticity and continuity of $\vec{\Psi}^{\pm}(x,y,\lambda)$.
$\vec{\Psi}^{\pm}(x,y,\lambda)$ satisfy the following
asymptotics for $|\lambda|\gg 1$,
\begin{equation}\label{e14}
    \vec{\Psi}^{\pm}(x,y,\lambda)=\vec{\eta}+\frac{1}{\lambda}
    \vec{U}(x,y)+\vec{O}(\frac{1}{\lambda^2}),
\end{equation}
where
\begin{equation*}
    \vec{\eta}=\left(
        \begin{gathered}
            \lambda+2u\\
            x+\lambda y+2uy
        \end{gathered}
    \right), \; \vec{U}(x,y)=\left(
        \begin{gathered}
            2\partial_x^{-1}u_y\\
            2\partial_x^{-1}(u_yy)+2\partial_x^{-1}u
        \end{gathered}
    \right).
\end{equation*}\par
The analytic Green's functions (\ref{e6})
exhibit the following asymptotics for $y\rightarrow\pm\infty$,
\begin{equation*}
    \begin{aligned}
        G^{\pm}(x-x',y-y',\lambda)\rightarrow\mp\frac{1}{2\pi i[\gamma-\gamma'\pm i\epsilon]},\; y\rightarrow +\infty,\\
        G^{\pm}(x-x',y-y',\lambda)\rightarrow\mp\frac{1}{2\pi i[\gamma-\gamma'\mp i\epsilon]}, \;y\rightarrow -\infty,
    \end{aligned}
\end{equation*}
where $\gamma=x+\lambda y$, $\gamma'=x'+\lambda y'$.
The asymptotics imply that
 the limits of $\vec{\Psi}^+$ and $\vec{\Psi}^-$ as $y\rightarrow+\infty$
are analytic  in the upper and lower halves
of complex $\gamma$ plane, respectively, while the
the limits of $\vec{\Psi}^+$ and $\vec{\Psi}^-$ as $y\rightarrow-\infty$
are analytic  in the lower and upper halves
of complex $\gamma$ plane, respectively.
\subsection{Scattering and spectral data}\label{subsection2}
In this subsection, based on the above Jost eigenfunctions and analytic eigenfunctions, we construct the scattering and spectral data that are used in the IST method from the initial data.\par
From the limit of $\vec{\varphi}_-$ as $y\rightarrow+\infty$,
one can define the following scattering vector $\vec{\sigma}$,
\begin{equation}\label{e7}
    \lim_{y\rightarrow +\infty}\vec{\varphi}_-(x,y,\lambda)=\vec{S}(\vec{\xi})=\vec{\xi}+\vec{\sigma}(\vec{\xi}).
\end{equation}
The scattering data $\vec{\sigma}(\vec{\xi})$ can be constructed from (\ref{e3}) and (\ref{e7}).\par
Since Lax pair (\ref{e33}) is made of  vector fields, the space of eigenfunctions is actually a ring, i.e., if $f_1$ and $f_2$ are two solutions of $\hat{L}_1f=0$, then an arbitrary differentiable function $F(f_1, f_2)$ is also a solution\cite{RN12}.
The Jost eigenfunctions $\vec{\varphi}_+$ and $\vec{\varphi}_-$
can form a basis, respectively, and the analytic eigenfunctions  $\vec{\Psi}^\pm$
can be defined by
\begin{align}
    &\vec{\Psi}^{+}(x,y,\lambda)=\vec{\mathcal{K}}_-^+(\vec{\varphi}_-(x,y,\lambda))=\vec{\mathcal{K}}_+^-(\vec{\varphi}_+(x,y,\lambda)),\label{e8}\\
    &\vec{\Psi}^{-}(x,y,\lambda)=\vec{\mathcal{K}}_-^-(\vec{\varphi}_-(x,y,\lambda))=\vec{\mathcal{K}}_+^+(\vec{\varphi}_+(x,y,\lambda)),\;\lambda\in\mathbb{R},\label{e9}
\end{align}
where
\begin{equation}\label{e8a}
    \vec{\mathcal{K}}_b^a(\vec{\zeta})=\vec{\zeta}+\vec{\chi}_b^a(\vec{\zeta}),
\end{equation}
the superscript $a$ and subscript $b$ can be $+$ or $-$.
$\vec{\mathcal{K}}_b^a$ is vector function with vector variables $\vec{\zeta}$. Specifically, $\vec{\zeta}$ take the values of $\vec{\varphi}_+$ or $\vec{\varphi}_-$ in (\ref{e8}) and (\ref{e9}).
\par
From (\ref{e8}) and (\ref{e9}), it follows that
\begin{equation*}
    \begin{aligned}
        \lim_{y\rightarrow-\infty}\vec{\Psi}^{\pm}(x,y,\lambda)=\vec{\xi}+\vec{\chi}_-^\pm(\vec{\xi}),\\
        \lim_{y\rightarrow+\infty}\vec{\Psi}^{\pm}(x,y,\lambda)=\vec{\xi}+\vec{\chi}_+^\mp(\vec{\xi}).
    \end{aligned}
\end{equation*}
The analyticity properties of $\vec{\Psi}^+$ and $\vec{\Psi}^-$ imply that $\vec{\chi}_+^-(\vec{\xi})$
and $\vec{\chi}_-^-(\vec{\xi})$ are analytic in \textit{Im} $\gamma>0$
decaying at $\gamma\rightarrow\infty$ like $O(\gamma^{-1})$,
while $\vec{\chi}_+^+(\vec{\xi})$ and $\vec{\chi}_-^+(\vec{\xi})$
are analytic in \textit{Im} $\gamma<0$
decaying at $\gamma\rightarrow\infty$ like $O(\gamma^{-1})$.\par
The above spectral vectors $\vec{\chi}_b^a$
can be constructed from the
scattering vector $\vec{\sigma}$ through some linear integral equations. Manakov and Santini have previously used  the Fourier transform to show this process\cite{RN14,RN15,RN16,RN17}. In addition, the RH problem with a shift can be used to investigate this problemin \cite{RN19}. The authors used the associated RH problems with a shift to construct spectral data for the Dunajski hierarchy. Taking the limit of the second equality of (\ref{e8}) as $y\rightarrow+\infty$, one obtains the following equations,
\begin{equation}\label{e10}
    \vec{\chi}_+^-(\vec{\xi})-\vec{\chi}_-^+(\vec{\xi}+\vec{\sigma}(\vec{\xi}))=\vec{\sigma}(\vec{\xi}), \;\lambda\in\mathbb{R}.
\end{equation}
In fact, they can be viewed as two
linear scalar RH problems in the variable $\vec{\xi}$, with the given shift
$\vec{\sigma}(\vec{\xi})$ for the unknown $\vec{\chi}_+^-$
and $\vec{\chi}_-^+$. Using relevant theory in \cite{RN20}, one can obtain
\begin{equation}\label{e10a}
    \begin{aligned}
        &\frac{1}{2}\chi_{-j}^+(\vec{\xi}+\vec{\sigma}(\vec{\xi}))+\frac{1}{2\pi i}
    \int_{\mathbb{R}}\frac{\chi_{-j}^+(\vec{\xi'})}{\gamma'-(\gamma+\sigma_2(\vec{\xi}))}\mathrm{d}\gamma'=0,\\
    &-\frac{1}{2}\chi_{+j}^-(\vec{\xi})+\frac{1}{2\pi i}
    \int_{\mathbb{R}}\frac{\chi_{+j}^-(\vec{\xi'})}{\gamma'-\gamma}\mathrm{d}\gamma'=0, \;j=1, 2.
    \end{aligned}
\end{equation}
Combining equations (\ref{e10}) and (\ref{e10a}), these two RH
problems with a shift are equivalent to the  linear Fredholm equations
\begin{equation}\label{e30}
    \chi_{+j}^-(\vec{\xi})+\frac{1}{2\pi i}\int_{\mathbb{R}}
    K(\vec{\xi},\vec{\xi}')\chi_{+j}^-(\vec{\xi}')\mathrm{d}\gamma'
    -M_j(\vec{\xi})=0,\;j=1,2,\;\lambda\in\mathbb{R},
\end{equation}
where
\begin{align*}
    &K(\vec{\xi},\vec{\xi}')=\frac{\partial(S(\vec{\xi}'))/\partial \gamma'}{S(\vec{\xi}')-S(\vec{\xi})}
    -\frac{1}{\gamma'-\gamma},\\
    &M_j(\vec{\xi})=\frac{1}{2}\sigma_j(\vec{\xi})+\frac{1}{2\pi i}\int_{\mathbb{R}}
    \frac{\partial(S(\vec{\xi}'))/\partial \gamma'}{S(\vec{\xi}')-S(\vec{\xi})}
    \sigma_j(\vec{\xi}')\mathrm{d}\gamma',\;j=1,2,\\
    &S(\vec{\xi})=\gamma+\sigma_2(\vec{\xi}), S(\vec{\xi}')=\gamma'+\sigma_2(\vec{\xi}'),\\
    &\vec{\xi}=(\lambda,\gamma)^T,\;\vec{\xi}'=(\lambda,\gamma')^T,
\end{align*}
and the integrals in (\ref{e30}) are understood in the sense of the principal value.\par
In addition, if the potential $u\in \mathbb{R}$, then the vector fields are real for
$\lambda\in \mathbb{R}$, and
\begin{equation*}
    \begin{aligned}
        &\vec{\varphi}_\pm\in \mathbb{R}^2,\; \overline{\vec{\Psi}^-}=\vec{\Psi}^+, \;\lambda\in\mathbb{R},\\ &
        \vec{\sigma}\in \mathbb{R}^2,\;\overline{\vec{\mathcal{K}}_a^-}=\vec{\mathcal{K}}_a^+,\;
        \overline{\vec{\chi}_a^-}=\vec{\chi}_a^+.
    \end{aligned}
\end{equation*}\par
From above process, we analyze the direct problem, obtaining the scattering data $\vec{\sigma}(\vec{\xi})$ and the spectral data $\vec{\chi}_+^-(\vec{\xi})$. Specifically, we construct
 (i) the scattering data $\vec{\sigma}(\vec{\xi})$ from
the initial data $u(x,y,0)$ and Jost eigenfunctions $\vec{\varphi}_-$, and (ii) the spectral data $\vec{\chi}_+^-(\vec{\xi})$  from solving the RH problems with a shift.

\subsection{Time evolution of the scattering and spectral data}
An important feature of integrable systems with Lax integrability is that the time variable usually appears in only one of the two Lax operators, indicating that the time evolution of these systems is reflected in that operator. In above direct problem, we only use the first operator $\hat{L}_1$ which characterizes spatial variations. In ths subsection, by using the second operator
$\hat{L}_2$, we investigate the time evolution of the scattering and spectral data.\par
 We denote these data containing $t$ by $\vec{\tilde{S}}(\zeta_1,\zeta_2,t)$, $\vec{\tilde{\mathcal{K}}}_b^a(\zeta_1,\zeta_2,t)$,
where
\begin{equation*}
    \vec{\tilde{S}}(\zeta_1,\zeta_2,0)=\vec{S}(\zeta_1,\zeta_2),\;
    \vec{\tilde{\mathcal{K}}}_b^a(\zeta_1,\zeta_2,0)=\vec{\mathcal{K}}_b^a(\zeta_1,\zeta_2).
\end{equation*}
These data can be described by the equations
\begin{equation}\label{e15}
    \begin{aligned}
        &(\partial_t+\frac{1}{2}\lambda^2\partial_{\zeta_2})f_1=0,\\
        &(\partial_t+\frac{1}{2}\lambda^2\partial_{\zeta_2})f_2=\frac{1}{2}f_1^2,
    \end{aligned}
\end{equation}
where $f_1(\zeta_1,\zeta_2,t)$ and $f_2(\zeta_1,\zeta_2,t)$ are the two components of
$\vec{f}(\zeta_1,\zeta_2,t)$.
The solutions $f_1$ and $f_2$  can be expressed as
\begin{equation}\label{e16}
    \begin{aligned}
        &f_1(\zeta_1,\zeta_2,t)=f_1(\zeta_1,\zeta_2-\frac{1}{2}t\zeta_1^2,0),\\
        &f_2(\zeta_1,\zeta_2,t)=f_2(\zeta_1,\zeta_2-\frac{1}{2}t\zeta_1^2,0)+\frac{1}{2}t(f_1(\zeta_1,\zeta_2-\frac{1}{2}t\zeta_1^2,0))^2.
    \end{aligned}
\end{equation}\par
In fact,
\begin{equation}\label{e39}
    \begin{aligned}
        &\phi_{-1}(x,y,t,\lambda)=\varphi_{-1}(x,y,t,\lambda),\\
        &\phi_{-2}(x,y,t,\lambda)=\varphi_{-2}(x,y,t,\lambda)-\frac{1}{2}t(\varphi_{-1}(x,y,t,\lambda))^2
    \end{aligned}
\end{equation}
are  common Jost eigenfunctions of $\hat{L}_1$, $\hat{L}_2$,
and they are a basis of the space of eigenfunctions.
The  limit of equation
$\hat{L}_2\phi_{-1}=0$ as $y\rightarrow+\infty$ yields $(\partial_t+\frac{1}{2}\lambda^2\partial_\gamma)\tilde{S}_1(\lambda,\gamma,t)=0$,
while the  limit of equation
$\hat{L}_2\phi_{-2}=0$ as $y\rightarrow+\infty$ yields $(\partial_t+\frac{1}{2}\lambda^2\partial_\gamma)(\tilde{S}_2(\lambda,\gamma,t)-\frac{1}{2}t(\tilde{S}_1(\lambda,\gamma,t))^2)=0$,
implying that $\tilde{S}_1$ and $\tilde{S}_2$ can be described by (\ref{e15}).\par
Similarly,
\begin{equation}\label{e17}
    \begin{aligned}
        &\pi^\pm_1(x,y,t,\lambda)=\Psi^\pm_1(x,y,t,\lambda),\\
        &\pi^\pm_2(x,y,t,\lambda)=\Psi^\pm_2(x,y,t,\lambda)-\frac{1}{2}t({\Psi^\pm_1}(x,y,t,\lambda))^2
    \end{aligned}
\end{equation}
are  common analytic eigenfunctions of $\hat{L}_1$ and $\hat{L}_2$.
$\phi_{-1}$ and $\phi_{-2}$ are a basis of the space of eigenfunctions, and
\begin{equation}\label{e49}
    \pi^\pm_1=\kappa^\pm_{-1}(\phi_{-1},\phi_{-2}),\; \pi^\pm_2=\kappa^\pm_{-2}(\phi_{-1},\phi_{-2}),\;\lambda\in\mathbb{R},
\end{equation}
where $\kappa^+_{-1}$ and $\kappa^+_{-2}$
are two components of $\vec{\kappa}^+_-$, while $\kappa^-_{-1}$ and $\kappa^-_{-2}$
are two components of $\vec{\kappa}^-_-$, and they don't depend explicitly on $t$.
At $t=0$,
\begin{equation}\label{e50}
        \vec{\tilde{\mathcal{K}}}_-^\pm(\varphi_{-1}(x,y,0,\lambda),\varphi_{-2}(x,y,0,\lambda),0)=\vec{\kappa}_-^\pm(\varphi_{-1}(x,y,0,\lambda),\varphi_{-2}(x,y,0,\lambda)),\;\lambda\in\mathbb{R}.
\end{equation}
From (\ref{e39}), (\ref{e49}) and (\ref{e50}), it follows that
\begin{equation}\label{e51}
    \vec{\pi}^\pm(x,y,t,\lambda)=\vec{\tilde{\mathcal{K}}}_-^\pm(\varphi_{-1}(x,y,t,\lambda),\varphi_{-2}(x,y,t,\lambda)-\frac{1}{2}t(\varphi_{-1}(x,y,t,\lambda))^2,0),\;\lambda\in\mathbb{R}.
\end{equation}
According to (\ref{e8}), (\ref{e9}), (\ref{e17}) and (\ref{e51}), one obtains
\begin{equation}
    \begin{aligned}
        &\tilde{\mathcal{K}}_{-1}^\pm(\varphi_{-1},\varphi_{-2},t)=\tilde{\mathcal{K}}_{-1}^\pm(\varphi_{-1},\varphi_{-2}-\frac{1}{2}t(\varphi_{-1})^2,0),\\
        &\tilde{\mathcal{K}}_{-2}^\pm(\varphi_{-1},\varphi_{-2},t)=
        \tilde{\mathcal{K}}_{-2}^\pm(\varphi_{-1},\varphi_{-2}-\frac{1}{2}t(\varphi_{-1})^2,0)+\frac{1}{2}t(\tilde{\mathcal{K}}_{-1}^\pm(\varphi_{-1},\varphi_{-2}-\frac{1}{2}t(\varphi_{-1})^2,0))^2,
    \end{aligned}
\end{equation}
where $\tilde{\mathcal{K}}_{-1}^\pm$ and $\tilde{\mathcal{K}}_{-2}^\pm$ two components of $\vec{\tilde{\mathcal{K}}}_-^\pm$.
 It implies that $\vec{\tilde{\mathcal{K}}}_-^\pm$ can be described by (\ref{e16}). Similarly, it holds for $\vec{\tilde{\mathcal{K}}}_+^\pm$.
\subsection{Inverse problem}\label{subsection3}
  In the investigation of the dispersionless integrable systems by using Manakov-Santini IST method, the inverse problem can be completed by solving an associated nonlinear RH problem. In this subsection, we concentrate on the construction of the inverse problem.\par
It follows from (\ref{e39}) and (\ref{e51}) that
\begin{align}
    &\vec{\pi}^{+}(x,y,t,\lambda)=\vec{\tilde{\mathcal{K}}}_-^+(\vec{\phi}_-(x,y,t,\lambda),0),\label{e8aa}\\
    &\vec{\pi}^{-}(x,y,t,\lambda)=\vec{\tilde{\mathcal{K}}}_-^-(\vec{\phi}_-(x,y,t,\lambda),0),\;\lambda\in\mathbb{R}.\label{e9aa}
\end{align}
From (\ref{e8aa}) and (\ref{e9aa}), one can construct the following nonlinear RH problem on the real axis,
\begin{equation}\label{e19a}
    \vec{\pi}^{+}(\lambda)=\vec{\mathcal{R}}(\vec{\pi}^{-}(\lambda))=\vec{\pi}^{-}(\lambda)+\vec{R}(\vec{\pi}^{-}(\lambda)), \;\lambda\in\mathbb{R},
\end{equation}
where $\vec{\pi}^{\pm}(\lambda)=(\pi^\pm_1(\lambda), \pi^\pm_2(\lambda))^T\in \mathbb{C}^2$, and the the RH data $\vec{\mathcal{R}}$ are constructed from the given spectral data $\vec{\tilde{\mathcal{K}}}_-^\pm$. The solutions $\vec{\pi}^{\pm}(\lambda)$
are analytic in the upper and lower
halves of the complex $\lambda$ plane,  respectively,
with the normalizations
\begin{equation}\label{e19aa}
    \vec{\psi}^{\pm}(\lambda)=\left(
        \begin{gathered}
            \lambda+2u\\
            -\frac{t}{2}\lambda^2+(y-2ut)\lambda+x-2(u^2+\partial^{-1}_xu_y)t+2uy
        \end{gathered}
    \right)+\vec{O}(\lambda^{-1}),\; |\lambda|\gg 1.
\end{equation}\par
Subtracting $\vec{\xi}$ from (\ref{e8aa}) and (\ref{e9aa}),
and applying the operators $\hat{P}_-$ and $\hat{P}_+$, defined by
\begin{equation}\label{e22}
    \hat{P}_\pm f(\lambda)=\pm\frac{1}{2\pi i}\int_{\mathbb{R}}\frac{f(\lambda')}{\lambda'-(\lambda\pm i \epsilon)}\mathrm{d}\lambda',\;\lambda\in\mathbb{R},
\end{equation}
to the resultant equalities, respectively,
one obtains
\begin{equation*}
    \begin{aligned}
        \frac{1}{2\pi i}P V \int_{\mathbb{R}}\frac{(\vec{\phi}_--\vec{\xi})}{\lambda'-\lambda}\mathrm{d}\lambda'+\frac{1}{2}(\vec{\phi}_--\vec{\xi})
        +\frac{1}{2\pi i}\int_{\mathbb{R}}\frac{\vec{\chi}_-^+(\vec{\phi}_-)}{\lambda'-(\lambda+ i \epsilon)}\mathrm{d}\lambda'=0,\\
        -\frac{1}{2\pi i}P V \int_{\mathbb{R}}\frac{(\vec{\phi}_--\vec{\xi})}{\lambda'-\lambda}\mathrm{d}\lambda'+\frac{1}{2}(\vec{\phi}_--\vec{\xi})
        -\frac{1}{2\pi i}\int_{\mathbb{R}}\frac{\vec{\chi}_-^-(\vec{\phi}_-)}{\lambda'-(\lambda- i \epsilon)}\mathrm{d}\lambda'=0,
    \end{aligned}
\end{equation*}
where $P V\int_{\mathbb{R}}$ represents the Cauchy principal value integral.
It follows that
\begin{equation}\label{e13}
    \vec{\phi}_-+\frac{1}{2\pi i}\int_{\mathbb{R}}\frac{\vec{\chi}_-^+(\vec{\phi}_-)}{\lambda'-(\lambda+ i \epsilon)}\mathrm{d}\lambda'
    -\frac{1}{2\pi i}\int_{\mathbb{R}}\frac{\vec{\chi}_-^-(\vec{\phi}_-)}{\lambda'-(\lambda- i \epsilon)}\mathrm{d}\lambda'=\vec{\xi}.
\end{equation}\par
With given spectral data,
 $\vec{\phi}_-$ can be reconstructed from (\ref{e13}). Then, the RH data $\vec{\mathcal{R}}$ is obtained from equations (\ref{e8aa})-(\ref{e19a}).
It follows from (\ref{e19aa}) that
\begin{equation}\label{e19b}
    \partial_x^{-1}u_y=\frac{1}{2}\lim_{\lambda\rightarrow\infty}\lambda(\psi^\pm_1(\lambda)-\lambda-2u).
\end{equation}
From equations (\ref{e19a}) and (\ref{e19b}), one can construct
 the potential $u$.
\subsection{Time evolution of the scattering and spectral data for an associated hierarchy}
  In fact, the integrable hierarchy formed by the infinite symmetries of  (\ref{e1}) has been established in \cite{RN21}. The hierarchy reflects the commutation between $\hat{L}_1$ and higher-order time operators, implying that the direct and inverse problems of the hierarchy are similar with above subsections.
    In this subsection, we concentrate on the time evolution of the scattering and spectral data related to higher-order time operators.\par
   In \cite{RN21}, the hamiltonian vector field Lax operators for the hierarchy is defined as
   \begin{equation}\label{e35}
       \hat{L}_n=\partial_{t_n}-\{H_n, \cdot\},\; n=1, 2, 3, \cdots,
   \end{equation}
   where
   \begin{equation*}
       H_n=\frac{(-1)^{n-1}}{n^2}(\Phi^n)_{\ge 0},
   \end{equation*}
   $()_{\ge 0}$ represents that the non-negative powers of
   $\lambda$ are retained in the expansion of a function with respect to
   $\lambda$,
   the symbol $\{,\}$ stands for the special Poisson bracket in two-dimensional phase space $(\lambda, x)$,
   \begin{equation*}
       \{f,g\}=\lambda\frac{\partial f}{\partial\lambda}\frac{\partial g}{\partial x}
       -\lambda\frac{\partial f}{\partial x}\frac{\partial g}{\partial \lambda},
   \end{equation*}
   and $\Phi$ is the solution of $\hat{L}_1f=0$ with the formal
   Laurent expansion
   \begin{equation*}
       \Phi=\lambda+\sum_{k\le 0}f_k\lambda^k.
   \end{equation*}\par
Similar to (\ref{e15}), the time dependence of the two components of the associated data
   $\vec{\tilde{S}}(\zeta_1,\zeta_2,t_n)$ and $\vec{\tilde{\mathcal{K}}}_b^a(\zeta_1,\zeta_2,t_n)$  can be described as
   \begin{equation}\label{e36}
       \begin{aligned}
           &(\partial_{t_n}+\frac{(-1)^n}{n}\lambda^n\partial_{\zeta_2})D_1=0,\\
           &(\partial_{t_n}+\frac{(-1)^n}{n}\lambda^n\partial_{\zeta_2})D_2=\frac{(-1)^n}{n}D_1^n,
       \end{aligned}
   \end{equation}
   whose explicit solutions can be expressed as
   \begin{equation*}
       \begin{aligned}
           &D_1(\zeta_1,\zeta_2,t_n)=D_1(\zeta_1,\zeta_2-\frac{(-1)^n}{n}t_n\zeta_1^n,0),\\
           &D_2(\zeta_1,\zeta_2,t_n)=D_2(\zeta_1,\zeta_2-\frac{(-1)^n}{n}t_n\zeta_1^n,0)+\frac{(-1)^n}{n}t_n(D_1(\zeta_1,\zeta_2-\frac{(-1)^n}{n}t_n\zeta_1^n,0))^n.
       \end{aligned}
   \end{equation*}\par
   Equations (\ref{e36})  follow from
   the common Jost eigenfunctions $\vec{\phi}_-(x,y,t_n,\lambda)$ and the common analytic eigenfunctions  $\vec{\pi}^\pm(x,y,t_n,\lambda)$ of
   both $\hat{L}_1$ and $\hat{L}_n$, $n\ge 2$, that are constructed from the Jost eigenfunctions $\vec{\varphi}_-(x,y,t_n,\lambda)$ and the analytic eigenfunctions  $\vec{\Psi}^\pm(x,y,t_n,\lambda)$ of $\hat{L}_1$, respectively, via the formulae
   \begin{equation}\label{e38}
       \begin{aligned}
           &\phi_{-1}(x,y,t_n,\lambda)=\varphi_{-1}(x,y,t_n,\lambda),\\
           &\phi_{-2}(x,y,t_n,\lambda)=\varphi_{-2}(x,y,t_n,\lambda)-\frac{(-1)^n}{n}t_n(\varphi_{-1}(x,y,t_n,\lambda))^n,\\
           &\pi_1^\pm(x,y,t_n,\lambda)=\Psi_1^\pm(x,y,t_n,\lambda),\\
           &\pi_2^\pm(x,y,t_n,\lambda)=\Psi_2^\pm(x,y,t_n,\lambda)-\frac{(-1)^n}{n}t_n(\Psi_1^\pm(x,y,t_n,\lambda))^n.
       \end{aligned}
   \end{equation}
\section{\sc \bf The longtime behaviour of the solutions}\label{sec3}
In the research of evolution systems, the longtime behaviour, particularly the asymptotic properties of the solutions as time tends to infinity, is often a significant focus. An associated nonlinear RH dressing has been presented in \cite{RN21}, and it can provide some integral equations to construct the longtime behaviour. In this section, we concentrate on the investigation of the longtime behaviour of the solutions.\par
In \cite{RN21}, the authors considered a nonlinear RH dressing with specific normalizations and a reality constraint, allowing the real potential $u$ to be reconstructed. The nonlinear RH dressing is defined as
\begin{equation}\label{e18}
    \vec{\pi}^+(\lambda)=\vec{\mathcal{R}}(\pi^{-}_1(\lambda),\pi^{-}_2(\lambda))=\vec{\pi}^-(\lambda)+\vec{R}(\pi^{-}_1(\lambda),\pi^{-}_2(\lambda)),\;\lambda\in\mathbb{R},
\end{equation}
where $\vec{\mathcal{R}}(\zeta_1,\zeta_2)
\in \mathbb{C}^2,\; (\zeta_1,\zeta_2)\in \mathbb{C}^2$ is the RH
data and the solutions $\vec{\pi}^{\pm}(\lambda)=(\pi^{\pm}_1,\pi^{\pm}_2)^T\in \mathbb{C}^2$
are analytic in the upper and lower halves of the complex $\lambda$ plane, respectively, with the normalizations
\begin{equation}\label{e19}
    \vec{\pi}^{\pm}(\lambda)=\left(
        \begin{gathered}
            \lambda+2u\\
            -\frac{t}{2}\lambda^2+(y-2ut)\lambda+x-2(u^2+\partial^{-1}_xu_y)t+2uy
        \end{gathered}
    \right)+\vec{O}(\lambda^{-1}),\; |\lambda|\gg 1,
\end{equation}
where $u$ satisfies
\begin{equation}\label{e20}
    \partial_x^{-1}u_y=\frac{1}{2}\lim_{\lambda\rightarrow\infty}\lambda(\pi^{\pm }_1(\lambda)-\lambda-2u).
\end{equation}\par
Supposing above inverse problem and its linearized problem are uniquely
solvable, the solutions of (\ref{e18}) are common eigenfunctions
of $\hat{L}_j$, $j=1,2$. Then the solutions  of (\ref{e1}) can be reconstructed from (\ref{e20}). In addition,
supposing the spectral data satisfies the reality constraint
\begin{equation}\label{e21}
    \vec{\mathcal{R}}(\overline{\vec{\mathcal{R}}(\overline{\vec{\zeta}})})=\vec{\zeta},
\end{equation}
where $\vec{\zeta}=(\zeta_1,\zeta_2)\in \mathbb{C}^2$, $u$ is real.\par
$\pi^{-}_1$ and $\pi^{-}_2$ can be characterized as
\begin{equation}\label{e23}
    \begin{aligned}
        \pi^{-}_1(\lambda)=&\;\lambda+2u+\frac{1}{2\pi i}\int_{\mathbb{R}}\frac{\mathrm{d}\lambda'}{\lambda'-(\lambda- i \epsilon)}R_1(\pi^{-}_1(\lambda'),\pi^{-}_2(\lambda')),\\
        \pi^{-}_2(\lambda)= &-\frac{t}{2}\lambda^2+(y-2ut)\lambda+x-2(u^2+\partial^{-1}_xu_y)t+2uy\\&+\frac{1}{2\pi i}\int_{\mathbb{R}}\frac{\mathrm{d}\lambda'}{\lambda'-(\lambda- i \epsilon)}R_2(\pi^{-}_1(\lambda'),\pi^{-}_2(\lambda')).
    \end{aligned}
\end{equation}
From (\ref{e20}), one can get the following integral equation,
\begin{equation}\label{e24}
    \partial_x^{-1}u_y=-\frac{1}{4\pi i}\int_{\mathbb{R}}\mathrm{d}\lambda R_1(\pi^{-}_1(\lambda),\pi^{-}_2(\lambda)).
\end{equation}
 In fact, by (\ref{e23}), equation (\ref{e24}) can be expressed as the following form,
\begin{equation}\label{e31}
    \partial_x^{-1}u_y=F(x-2(u^2+\partial^{-1}_xu_y)t+2uy,\;y-2ut,\;t),
\end{equation}
where
\begin{equation}\label{e32}
    F(X, Y, t)=-\frac{1}{4\pi i}\int_{\mathbb{R}}\mathrm{d}\lambda R_1(\pi^{-}_1(\lambda;X,Y,t),\pi^{-}_2(\lambda;X,Y,t)).
\end{equation}\par
Consider the longtime behaviour in the space-time region
\begin{equation}\label{e25}
    \begin{aligned}
        &x=\tilde{x}+v_1t,\; y=\tilde{y}+v_2t,\\
        &\tilde{x}-2(u^2+\partial^{-1}_xu_y)t, \;\tilde{y}-2ut,\; v_1,\; v_2=O(1),\quad t\gg 1.
    \end{aligned}
\end{equation}
We define
\begin{equation}\label{e26}
    \begin{gathered}
        \phi_j(\lambda)=\frac{1}{2\pi i}\int_{\mathbb{R}}\frac{\mathrm{d}\lambda'}{\lambda'-(\lambda- i \epsilon)}R_j(\pi^{-}_1(\lambda'),\pi^{-}_2(\lambda')),\;j=1,2.
    \end{gathered}
\end{equation}
Then $\pi^{-}_1$ and $\pi^{-}_2$ can be rewritten as the following form,
\begin{equation}\label{e27}
    \begin{gathered}
        \pi^{-}_1(\lambda)=\phi_1(\lambda)+\lambda+2u,\\
        \pi^{-}_2(\lambda)=\phi_2(\lambda)-\frac{t}{2}\lambda^2+(y-2ut)\lambda+x-2(u^2+\partial^{-1}_xu_y)t+2uy.
    \end{gathered}
\end{equation}
Substituting (\ref{e27}) into (\ref{e26}) and then according to (\ref{e25}), one obtains
\begin{equation}\label{e28}
    \begin{gathered}
        \phi_j(\lambda)=\frac{1}{2\pi i}\int_{\mathbb{R}}\frac{\mathrm{d}\lambda'}{\lambda'-(\lambda- i \epsilon)}R_j
        (\lambda'+2u+\phi_1(\lambda'),-\frac{1}{2}(\lambda'-\lambda_+)(\lambda'-\lambda_-)t
        \\+(\tilde{y}-2ut)(\lambda'-v_2)+\tilde{x}-2(u^2+\partial^{-1}_xu_y)t+\tilde{y}(2u+v_2)+\phi_2(\lambda')),
        \;j=1, 2,
    \end{gathered}
\end{equation}
where
\begin{equation*}
    \lambda_\pm=v_2\pm\sqrt{2v_1+v_2^2}.
\end{equation*}
The fast decay is partially contrasted if $\lambda_+=\lambda_-$ for $t\gg 1$, and it is equivalent to the parabola
\begin{equation}\label{e34}
    x+\frac{(y-\tilde{y})^2}{2t}=\tilde{x} \quad\quad(2v_1+v_2^2=0)
\end{equation}
of the $(x, y)$-plane.\par
Thus, on such a parabola the integral equations read as
\begin{equation*}
    \begin{gathered}
        \phi_j(\lambda)=\frac{1}{2\pi i}\int_{\mathbb{R}}\frac{\mathrm{d}\lambda'}{\lambda'-(\lambda- i \epsilon)}R_j
        (\lambda'+2u+\phi_1(\lambda'),-\frac{1}{2}(\lambda'-v_2)^2t+
        (\tilde{y}-2ut)(\lambda'-v_2)\\+\tilde{x}-2(u^2+\partial^{-1}_xu_y)t+\tilde{y}(2u+v_2)+\phi_2(\lambda')),
        \;j=1, 2.
    \end{gathered}
\end{equation*}
Notice that the main contribution to the integral equations occurs when
$\lambda'\sim v_2$, thus one can make the change of variables
$\lambda'=v_2+\frac{\mu'}{\sqrt{t}}$, obtaining
\begin{equation}\label{e29}
    \begin{gathered}
        \phi_j(\lambda)=\frac{1}{2\pi i\sqrt{t}}\int_{\mathbb{R}}\frac{\mathrm{d}\mu'}{\frac{\mu'}{\sqrt{t}}-(\lambda-v_2-i \epsilon)}R_j
        (v_2+\frac{\mu'}{\sqrt{t}}+2u+\phi_1(v_2+\frac{\mu'}{\sqrt{t}}),-\frac{1}{2}\mu'^2\\+
        (\tilde{y}-2ut)\frac{\mu'}{\sqrt{t}}+\tilde{x}-2(u^2+\partial^{-1}_xu_y)t+\tilde{y}(2u+v_2)+\phi_2(v_2+\frac{\mu'}{\sqrt{t}})),
        \;j=1, 2.
    \end{gathered}
\end{equation}
If $|\lambda-v_2|\gg t^{-\frac{1}{2}}$, according to (\ref{e29}), one obtains
\begin{equation*}
    \begin{gathered}
        \phi_j(\lambda)\sim-\frac{1}{2\pi i\sqrt{t}(\lambda-v_2-i \epsilon)}\int_{\mathbb{R}}\mathrm{d}\mu'R_j
        (v_2+2u+\phi_1(v_2+\frac{\mu'}{\sqrt{t}}),-\frac{1}{2}\mu'^2\\+\tilde{x}-2(u^2+\partial^{-1}_xu_y)t+\tilde{y}(2u+v_2)+\phi_2(v_2+\frac{\mu'}{\sqrt{t}})),
        \;j=1, 2.
    \end{gathered}
\end{equation*}
It implies that $\phi_j(\lambda)=O(t^{-\frac{1}{2}}), j=1, 2$, and by (\ref{e24}), that
\begin{equation*}
    \begin{gathered}
        \partial_x^{-1}u_y=-\frac{1}{4\pi i\sqrt{t}}\int_{\mathbb{R}}\mathrm{d}\mu R_1
        (v_2+2u+\phi_1(v_2+\frac{\mu}{\sqrt{t}}),-\frac{1}{2}\mu^2+\tilde{x}\\-2(u^2+\partial^{-1}_xu_y)t+\tilde{y}(2u+v_2)+\phi_2(v_2+\frac{\mu}{\sqrt{t}}))+o(\frac{1}{\sqrt{t}}).
    \end{gathered}
\end{equation*}
Similarly, if $\lambda-v_2=\frac{\mu}{\sqrt{t}}$ , according to (\ref{e29}),
one obtains
\begin{equation*}
    \begin{gathered}
        \phi_j(v_2+\frac{\mu}{\sqrt{t}})\sim \frac{1}{2\pi i}\int_{\mathbb{R}}\frac{\mathrm{d}\mu'}{\mu'-(\mu-i \epsilon)}R_j
        (v_2+2u+\phi_1(v_2+\frac{\mu'}{\sqrt{t}}),-\frac{1}{2}\mu'^2\\+\tilde{x}-2(u^2+\partial^{-1}_xu_y)t+\tilde{y}(2u+v_2)+\phi_2(v_2+\frac{\mu'}{\sqrt{t}})),
        \;j=1, 2.
    \end{gathered}
\end{equation*}
It implies that $\phi_j(\lambda)=O(1), \;j=1, 2.$
Therefore, one must take $\phi_j$ of $R_j$ into consideration,
implying that these integral equations remain nonlinear in the longtime regime.\par
The conclusions derived from above process can be summarized as  the following theorem.
\begin{thm}
In the space-time region
\begin{equation*}
    \begin{aligned}
        &x=\tilde{x}+v_1t,\; y=\tilde{y}+v_2t,\;x+\frac{(y-\tilde{y})^2}{2t}=\tilde{x},\\
        &\tilde{x}-2(u^2+\partial^{-1}_xu_y)t, \;\tilde{y}-2ut,\; v_1,\; v_2=O(1),\quad t\gg 1,
    \end{aligned}
\end{equation*}
the longtime behaviour of the solutions reads
\begin{equation}\label{e37}
    w=\frac{1}{\sqrt{t}}G(\frac{y-\tilde{y}}{t}+2u,\;x+\frac{(y-\tilde{y})^2}{2t}-2(u^2+\partial^{-1}_xu_y)t
    +\tilde{y}(2u+\frac{y-\tilde{y}}{t}))+o(\frac{1}{\sqrt{t}}),
\end{equation}
where
\begin{align*}
    &w_x(x,y,t)=u_y(x,y,t),\\
    &G(\alpha,\beta)=-\frac{1}{4\pi i}\int_{\mathbb{R}}\mathrm{d}\mu
    R_1(\alpha+a_1(\mu;\alpha,\beta),-\frac{1}{2}\mu^2+\beta+a_2(\mu;\alpha,\beta)),
\end{align*}
$R_1$ is the first component of the RH data $\vec{R}$, and $a_1(\mu;\alpha,\beta)$, $a_2(\mu;\alpha,\beta)$ are unique solutions of the following nonlinear
integral equations, respectively,
\begin{equation*}
    \begin{gathered}
        a_j(\mu;\alpha,\beta)=\frac{1}{2\pi i}\int_{\mathbb{R}}\frac{\mathrm{d}\mu'}{\mu'-(\mu- i \epsilon)}
        R_j(\alpha+a_1(\mu';\alpha,\beta),-\frac{1}{2}\mu'^2+\beta+a_2(\mu';\alpha,\beta)),\\
        \;j=1, 2.
    \end{gathered}
\end{equation*}
\end{thm}
The longtime behaviour of $w$ is influenced by both $u$ and the nonlinearity
 of $G$, and it is possible to obtain  (\ref{e37}). In addition, $G$ can be determined by the RH data
 constructed in Cauchy problem. The presence of $u$ in $G$ implies that there might occur a wave breaking,
and our future work is to study the possible wave breaking.
\section{\sc \bf The implicit solutions}
In this section, we present two examples of  the nonlinear RH dressing referred to in section \ref{sec3}, and construct the corresponding  implicit solutions of (\ref{e1}).\par
\textbf{Example 1}. Suppose that the components of the RH data in (\ref{e18}) are
\begin{equation*}
    \mathcal{R}_j(\zeta_1,\zeta_2)=\zeta_je^{i(-1)^{j+1}f_1(\zeta_1\zeta_2)},\;j=1,2,
\end{equation*}
where $f_1$ is an arbitrary real function of a single variable. Thus, the nonlinear RH problem becomes
\begin{equation}\label{e41}
    \begin{aligned}
        &\pi^{+}_1=\pi^{-}_1e^{if_1(\pi^{-}_1\pi^{-}_2)},\\
        &\pi^{+}_2=\pi^{-}_2e^{-if_1(\pi^{-}_1\pi^{-}_2)},\;\lambda\in\mathbb{R}.
    \end{aligned}
\end{equation}
From the associated theory of solvable vector nonlinear RH problems developed in \cite{RN22}, the problem satisfies the following properties.\par
(i) The reality constraint (\ref{e21}) holds.\par
(ii) $\pi^{+}_1\pi^{+}_2=\pi^{-}_1\pi^{-}_2$. Therefore, $\pi^{+}_1\pi^{+}_2$ is a polynomial in $\lambda$, and we denote it by $W_1(\lambda)$. Then, from (\ref{e19}),
\begin{equation}\label{e42}
    \begin{aligned}
        W_1(\lambda)=&-\frac{t}{2}\lambda^3+(y-3ut)\lambda^2+(x-3t\partial_x^{-1}u_y+4uy-6u^2t)\lambda-3t\partial_x^{-1}u_{yy}-8ut\partial_x^{-1}u_y\\
        &+4y\partial_x^{-1}u_y+2t\partial_x^{-1}(u_x\partial_x^{-1}u_y)-4t\partial_x^{-1}(uu_y)+2\partial_x^{-1}u+2ux-4u^3t+4u^2y.
    \end{aligned}
\end{equation}
The problem (\ref{e41}) becomes
\begin{equation}\label{e43}
    \pi^{+}_1=\pi^{-}_1e^{if_1(W_1(\lambda))},\; \pi^{+}_2=\pi^{-}_2e^{-if_1(W_1(\lambda))},\;\lambda\in\mathbb{R}.
\end{equation}\par
(iii) Using the operators (\ref{e22}), from (\ref{e43}), one obtains
\begin{equation}
    \pi^{+}_je^{i(-1)^jf^+_1(\lambda)}=\pi^{-}_je^{i(-1)^jf^-_1(\lambda)},\;j=1,2,\;\lambda\in\mathbb{R},
\end{equation}
where
\begin{equation}\label{e44}
    f^\pm_1(\lambda)=\frac{1}{2\pi i}\int_{\mathbb{R}}\frac{f_1(W_1(\lambda'))}{\lambda'-(\lambda\pm i \epsilon)}\mathrm{d}\lambda',\;\lambda\in\mathbb{R},
\end{equation}
implying that $\pi^{+}_je^{i(-1)^jf^+_1(\lambda)}, \;j=1,2$, are also polynomials in $\lambda$. For the convenience of subsequent writing, we introduce the notation
\begin{equation}\label{e45}
    <\lambda^nf_1>=\frac{1}{2\pi}\int_{\mathbb{R}}\lambda^nf_1(W_1(\lambda))\mathrm{d}\lambda,\;n=0,1,2,\cdots,\;\lambda\in\mathbb{R}.
\end{equation}\par
One can expand $f^+_1(\lambda)$ and $\pi^{+}_je^{i(-1)^jf^+_1(\lambda)}$ in powers of $\lambda$, obtaining the positive power expansion
\begin{equation}\label{e46}
    \begin{aligned}
        &\pi^{+}_1e^{-if^+_1(\lambda)}=\pi^{-}_1e^{-if^-_1(\lambda)}=\lambda+2u+<f_1> ,\\
        &\pi^{+}_2e^{if^+_1(\lambda)}=\pi^{-}_2e^{if^-_1(\lambda)}=-\frac{t}{2}\lambda^2+(\frac{t<f_1>}{2}+y-2ut)\lambda\\ &\quad\quad-(y-2ut+\frac{t}{4})<f_1> +\frac{t<\lambda f_1>}{2}+x-2(u^2+\partial^{-1}_xu_y)t+2uy.
    \end{aligned}
\end{equation}
Therefore, from (\ref{e46}), one obtains the following explicit solutions of the RH problem (\ref{e41}),
\begin{equation*}
    \begin{aligned}
        \pi^\pm_1&=(\lambda+2u+<f_1>)e^{if^\pm_1(\lambda)},\\
        \pi^\pm_2&=(-\frac{t}{2}\lambda^2+(\frac{t<f_1>}{2}+y-2ut)\lambda-(y-2ut+\frac{t}{4})<f_1>\\ &\quad\quad +\frac{t<\lambda f_1>}{2}+x-2(u^2+\partial^{-1}_xu_y)t+2uy)e^{-if^\pm_1(\lambda)}
    \end{aligned}
\end{equation*}
The coefficient of the $\lambda^{-1}$ term of (\ref{e46}) must be zero, and one can obtain the following implicit solution of (\ref{e1}),
\begin{equation*}
    \partial_x^{-1}u_y=-u<f_1>-\frac{<\lambda f_1>}{2}-\frac{<f_1>^2}{4}.
\end{equation*}\par
\textbf{Example 2}. Suppose that the components of the RH data in (\ref{e18}) are
\begin{equation*}
    \begin{aligned}
        &\mathcal{R}_1(\zeta_1,\zeta_2)=\zeta_1+if_2(c(\zeta_1)^n+\zeta_2),\\
        &\mathcal{R}_2(\zeta_1,\zeta_2)=\zeta_2+c(\zeta_1)^n-c(\zeta_1+if_2(c(\zeta_1)^n+\zeta_2))^n,
    \end{aligned}
\end{equation*}
where $n$ is a positive integer, $c$ is a real parameter, and $f_2$ is an arbitrary real function of a single variable. Thus, the nonlinear RH problem becomes
\begin{equation}\label{e61}
    \begin{aligned}
        &\pi^{+}_1=\pi^{-}_1+if_2(c(\pi^{-}_1)^n+\pi^{-}_2),\\
        &\pi^{+}_2=\pi^{-}_2+c(\pi^{-}_1)^n-c(\pi^{-}_1+if_2(c(\pi^{-}_1)^n+\pi^{-}_2))^n,\;\lambda\in\mathbb{R}.
    \end{aligned}
\end{equation}\par
They satisfy the reality constraint (\ref{e21}), and from (\ref{e61}),
$c(\pi^{+}_1)^n+\pi^{+}_2=c(\pi^{-}_1)^n+\pi^{-}_2$. Therefore, $c(\pi^{+}_1)^n+\pi^{+}_2$ is a polynomial in $\lambda$, and we denote it by $W_3(\lambda)$. Then, from (\ref{e19}),
\begin{equation}\label{e62}
    \begin{aligned}
        W_2(\lambda)=c((\pi^{+}_1)^n)_{\ge 0}-\frac{t}{2}\lambda^2+(y-2ut)\lambda+x-2(u^2+\partial^{-1}_xu_y)t+2uy,
    \end{aligned}
\end{equation}
where $()_{\ge 0}$ represents  the non-negative powers of $\lambda$ terms.
Then the first equation of (\ref{e61}) becomes
\begin{equation}\label{e63}
    \begin{aligned}
        \pi^{+}_1-if_2^+(\lambda)=\pi^{-}_1-if_2^-(\lambda),\;\lambda\in\mathbb{R},
    \end{aligned}
\end{equation}
where
\begin{equation}\label{e64}
    f_2^\pm(\lambda)=\frac{1}{2\pi i}\int_{\mathbb{R}}\frac{f_2(W_2(\lambda'))}{\lambda'-(\lambda\pm i \epsilon)}\mathrm{d}\lambda',\;\lambda\in\mathbb{R}.
\end{equation}
We denote
\begin{equation}\label{e65}
    <\lambda^nf_2>=\frac{1}{2\pi}\int_{\mathbb{R}}\lambda^nf_2(W_2(\lambda))\mathrm{d}\lambda,\;n=0,1,2,\cdots,\;\lambda\in\mathbb{R}.
\end{equation}
From equations (\ref{e62})-(\ref{e65}), one obtains the following explicit solutions of the RH problem (\ref{e61}),
\begin{equation}\label{e66}
    \begin{aligned}
        &\pi^{\pm}_1=\lambda+2u+if_2^\pm(\lambda),\\
        &\pi^{\pm}_2=W_2(\lambda)-c(\pi^{\pm}_1)^n.
    \end{aligned}
\end{equation}
Comparing the coefficients of the $\lambda$ terms on both sides of the first equation of (\ref{e66}), one can obtain the following implicit solution of (\ref{e1}),
\begin{equation*}
    \partial_x^{-1}u_y=-\frac{<f_2>}{2}.
\end{equation*}

\textbf{Acknowledgements:}
This work is supported by the National Natural Science Foundation of China under Grant Nos. 12271136, 12171133, 12171132.

% \nocite{*}
% \bibliographystyle{mystyle}
% \bibliography{myref}

\end{document}